%% file: main.tex
\documentclass[9pt,twocolumn,twoside]{pnas-new}
\setboolean{displaywatermark}{false}

\templatetype{pnasresearcharticle} 
\usepackage{fontawesome}

\newcommand{\given}{\,|\,}
\newcommand{\simbig}{{\sc SimBIG}}

\newcommand{\bfi}[1]{\textbf{\textit{#1}}}

\let\oldAA\AA
\renewcommand{\AA}{\text{\normalfont\oldAA}}

\newcommand{\btheta}{\boldsymbol{\theta}}
\newcommand{\bphi}{\boldsymbol{\phi}}

\title{\simbig: A Forward Modeling Approach To Analyzing Galaxy Clustering}

\author[a,1]{ChangHoon Hahn}
\author[b]{Michael Eickenberg}
\author[c]{Shirley Ho}
\author[d,e]{Jiamin Hou}
\author[f, g, c]{Pablo Lemos}
\author[h,i]{Elena Massara}
\author[b, c]{Chirag Modi}
\author[j]{Azadeh Moradinezhad Dizgah}
\author[b]{Bruno R\'egaldo-Saint Blancard}
\author[j]{Muntazir M. Abidi}

\affil[a]{Department of Astrophysical Sciences, Princeton University, Princeton NJ 08544, USA}
\affil[b]{Center for Computational Mathematics, Flatiron Institute, 162 5th Ave, New York, NY 10010, USA}
\affil[c]{Center for Computational Astrophysics, Flatiron Institute, 162 5th Ave 5th floor, New York, NY 10010, USA}
\affil[d]{Department of Astronomy, University of Florida, 211 Bryant Space Science Center, Gainesville, FL 32611, USA}
\affil[e]{Max-Planck-Institut f\"ur Extraterrestrische Physik, Postfach 1312, Giessenbachstrasse 1, 85748 Garching bei M\"unchen, Germany}
\affil[f]{Department of Physics, Universit\'{e} de Montr\'{e}al, Montr\'{e}al, 1375 Avenue Th\'{e}r\`{e}se-Lavoie-Roux, QC H2V 0B3, Canada}
\affil[g]{Mila - Quebec Artificial Intelligence Institute, Montr\'{e}al, 6666 Rue Saint-Urbain, QC H2S 3H1, Canada}
\affil[h]{Waterloo Centre for Astrophysics, University of Waterloo, 200 University Ave W, Waterloo, ON N2L 3G1, Canada}
\affil[i]{Department of Physics and Astronomy, University of Waterloo, 200 University Ave W, Waterloo, ON N2L 3G1, Canada}
\affil[j]{D\'epartement de Physique Th\'eorique, Universit\'e de Gen\`eve, 24 quai Ernest Ansermet, 1211 Gen\`eve 4, Switzerland}

\leadauthor{Hahn} 

\significancestatement{
The three-dimensional spatial distribution of galaxies encodes key cosmological 
information on the nature of dark energy and the contents of the Universe. 
Current analyses of the statistical clustering of galaxies successfully extract
information on large scales that are well described by analytic models.
They, however, struggle on smaller, non-linear, scales. 
Here we present \simbig, a new approach to galaxy clustering analyses that 
can extract information on non-linear regimes by exploiting high-fidelity 
simulations and inference based on machine learning. 
To demonstrate its advantages, we apply \simbig~to 
109,636 galaxies of the BOSS survey and analyze a standard summary statistic 
of the galaxy distribution. 
Our constraints are consistent with previous works and substantially improve 
their precision on select cosmological parameters.
}

\authorcontributions{
    C.H., M.E., S.H., P.L., E.M., A.M.D., B.R.B., and M.A. designed research; 
    C.H. performed research; 
    C.H., M.E., S.H., J.H., P.L., E.M., A.M.D., and B.R.B. contributed new reagents/analytic tools; 
    C.H., M.E., S.H., J.H., P.L., E.M., C.M., A.M.D., and B.R.B. analyzed data; and 
    C.H., M.E., S.H., J.H., P.L., E.M., C.M., A.M.D., and B.R.B. wrote the paper.}
\authordeclaration{The authors declare no conflict of interest.}
\correspondingauthor{\textsuperscript{1}changhoon.hahn@princeton.edu}


\begin{abstract}
    We present the first-ever cosmological constraints from a simulation-based 
    inference (SBI) analysis of galaxy clustering
    from the new \simbig~forward modeling framework.
    \simbig~leverages the predictive power of high-fidelity simulations 
    and provides an inference framework that can extract cosmological 
    information on small non-linear scales, inaccessible with standard 
    analyses. 
    In this work, we apply \simbig~to the BOSS CMASS galaxy sample and analyze 
    the power spectrum, $P_\ell(k)$, to  $k_{\rm max}=0.5\,h/{\rm Mpc}$. 
    We construct 20,000 simulated galaxy samples using our forward model, which is 
    based on high-resolution {\sc Quijote} $N$-body simulations and includes detailed 
    survey realism for a more complete treatment of observational systematics. 
    We then conduct SBI by training normalizing flows using the simulated samples 
    and infer the posterior distribution of $\Lambda$CDM cosmological parameters:
    $\Omega_m, \Omega_b, h, n_s, \sigma_8$.
    We derive significant constraints on $\Omega_m$ and $\sigma_8$, which are 
    consistent with previous works. 
    Our constraints on $\sigma_8$ are 27\% more precise than standard analyses.  
    This improvement is equivalent to the statistical gain expected from analyzing a 
    galaxy sample that is $\sim$60\% larger than CMASS with standard methods. 
    It results from additional cosmological information on non-linear scales beyond the
    limit of current analytic models, $k > 0.25\,h/{\rm Mpc}$.  
    While we focus on $P_\ell$ in this work for validation and comparison to the
    literature, \simbig~provides a framework for analyzing galaxy clustering using 
    any summary statistic. 
    We expect further improvements on cosmological constraints from subsequent 
    \simbig~analyses of summary statistics beyond $P_\ell$.
\end{abstract}

\dates{This manuscript was compiled on \today}
\doi{\url{www.pnas.org/cgi/doi/10.1073/pnas.XXXXXXXXXX}}

\begin{document}

\maketitle
\thispagestyle{firststyle}
\ifthenelse{\boolean{shortarticle}}{\ifthenelse{\boolean{singlecolumn}}{\abscontentformatted}{\abscontent}}{}


\input{intro}
\input{sbi}
\input{results}
\input{conclusion}

\matmethods{
\subsection*{Observations: SDSS-III BOSS} \label{sec:obs}
In this work, we analyze observations from the Sloan Digital Sky Survey
SDSS-III~\citep{eisenstein2011, dawson2013} Baryon Oscillation Spectroscopic
Survey (BOSS) Data Release 12. 
In particular, we use the CMASS galaxy sample, which selects high stellar mass
Luminous Red Galaxies (LRGs) over the redshift $0.43 < z <
0.7$~\citep{reid2016}.
We restrict our analysis to CMASS galaxies in the Southern Galactic Cap (SGC) 
and impose a redshift cut of $0.45 < z < 0.6$ and the following angular cuts: 
${\rm Dec} > -6$ and $-25 < {\rm RA} < 28$.
In upper panels in Fig.~\ref{fig:fm}, we present the three-dimensional 
distributions of our CMASS SGC galaxy sample at three different viewing 
angles. 
We also present the angular distribution of the sample in Fig.~\ref{fig:fm_demo_ang}.
In total, our CMASS SGC galaxy sample contains 109,636 galaxies.
}
\showmatmethods{} 

\acknow{
It is a pleasure to thank Peter Melchior, Uro{\u s}~Seljak, David Spergel, Licia Verde, and Benjamin D. Wandelt for valuable discussions. 
We also thank Mikhail M. Ivanov and Yosuke Kobayashi for providing us with the posteriors used in the comparison.  
This work was supported by the AI Accelerator program of the Schmidt Futures Foundation. 
This work was also supported by NASA ROSES grant 12-EUCLID12-0004 NASA grant 15-WFIRST15-0008. 
JH has received funding from the European Union’s Horizon 2020 research and innovation program under the Marie Sk\l{}odowska-Curie grant agreement No 101025187. 
AMD acknowledges funding from Tomalla Foundation for Research in Gravity and Boninchi Foundation. 
}

\showacknow{} 

\bibliography{simbig}

\end{document}

%% file: intro.tex
\dropcap{T}he three-dimensional spatial distribution of galaxies provides key 
cosmological information that can be used to constrain the nature of dark matter 
and dark energy and measure the contents of the Universe. 
The next generation spectroscopic galaxy surveys, conducted using the 
Dark Energy Spectroscopic Instrument~\citep[DESI;][]{desicollaboration2016, desicollaboration2016a, abareshi2022}, 
Subaru Prime Focus Spectrograph~\citep[PFS;][]{takada2014, tamura2016}, 
the ESA {\em Euclid} satellite mission~\citep{laureijs2011}, and the
Nancy Gracy Roman Space Telescope~\citep[Roman;][]{spergel2015, wang2022a}, 
will probe galaxies over unprecedented cosmic volumes out to $z\sim3$,
over 10 Gyrs of cosmic history.  
Combined with other cosmological probes, they will provide the most stringent 
tests of the standard $\Lambda$CDM cosmological model and potentially lead to 
discoveries of new physics.

In current analyses, the power spectrum is used as the primary measurement of 
galaxy clustering~\citep[\emph{e.g.}][]{beutler2017, ivanov2020, kobayashi2021}. 
Furthermore, the analyses are limited to large, linear scales where the impact of non-linear 
structure formation is small. 
These restrictions result from the fact that standard analyses use analytic 
models based on perturbation theory (PT) of large-scale 
structure~\citep[see][for a review]{bernardeau2002,desjacques2016}.
PT struggles to accurately model scales beyond quasi-linear scales, especially 
for higher-order clustering statistics (\emph{e.g.} bispectrum).
In the \cite{philcox2021} PT analysis, for instance, the authors restrict the power 
spectrum to $k < 0.2\,h/{\rm Mpc}$ and the bispectrum to $k < 0.08\,h/{\rm Mpc}$. 
In a recent PT analysis, \cite{damico2022} analyzes the bispectrum to 
$k < 0.23\,h/{\rm Mpc}$; however, they require 33 extra parameters for the 
theoretical consistency of their model.
PT also cannot be used to model various newly proposed summary
statistics~\citep[\emph{e.g.}][]{naidoo2022} or to exploit the full galaxy 
distribution at the field level. 

Another major challenge for current analyses is accurately accounting for
observational systematics. 
Observations suffer from imperfections in \emph{e.g.} targeting, imaging, 
completeness that can significantly impact the analysis~\citep[][]{ross2012, ross2017}.
Current analyses account for these effects by applying correction weights to
the galaxies. 
Fiber collisions, for example, prevent galaxy surveys that use fiber-fed spectrographs
(\emph{e.g.} DESI, PFS)
from successfully measuring redshifts from galaxies within some angular scale of 
one another on the focal plane.
This significantly biases the power spectrum by more than the amplitude 
of cosmic variance on scales smaller than 
$k > 0.1\,h/{\rm Mpc}$~\citep{guo2012, hahn2017a, bianchi2018}.
To correct for this effect, the weights of the ``collided'' galaxies
missed by survey are assigned to their nearest angular 
neighbors~\citep{zehavi2002, anderson2014}. 
Even for current analyses, these correction weights do not sufficiently correct 
the measured power spectrum~\cite{hahn2017a}.
Furthermore, they are only designed and demonstrated for the power spectrum.

Meanwhile, additional cosmological information is available on non-linear scales 
and in higher-order statistics. 
Recent studies have accurately quantified the information content in these regimes
using large suites of simulations. 
\cite{hahn2020} and \cite{hahn2021a} used the {\sc Quijote} suite of simulations 
to demonstrate that constraints on cosmological parameters, 
$\Omega_m, \Omega_b, h, n_s, \sigma_8$, improve by a factor of $\sim$2 by
including non-linear scales ($0.2 < k < 0.5\,h/{\rm Mpc}$) in power spectrum analyses. 
Even more improvement comes from including higher-order clustering information 
in the bispectrum.
Similar forecasts for other summary statistics, \emph{e.g.} marked power 
spectrum~\citep{massara2020, massara2022}, reconstructed power spectrum~\citep{wang2022}, 
skew spectra~\citep{hou2022}, wavelet statistics~\citep{eickenberg2022}, find
consistent improvements from including non-linear scales and higher-order clustering. 
Despite the growing evidence of the significant constraining power available in
non-linear scales and higher-order statistics, it cannot be exploited by 
standard methods with PT. 

Robustly exploiting non-linear and non-Gaussian cosmological information requires
a framework that can both accurately model non-linear structure formation and
account for detailed observational systematics. 
In this work, we present SIMulation-Based Inference of Galaxies (\simbig), a framework
for analyzing galaxy clustering that achieves these requirements by using a forward 
modeling approach. 
Instead of analyzing galaxy clustering using analytic models, a forward model approach
uses simulations that model the full details of the observations. 

In \simbig, our forward model is based on cosmological $N$-body simulations that 
accurately models non-linear structure formation.  
We also use the halo occupation framework, which provides a compact and flexible 
prescription for connecting the galaxy distribution to the dark matter distribution.
Our forward model also takes advantage of the fact that many observational 
systematics can be more easily included in 
simulations~\citep[\emph{e.g.}][]{rodriguez-torres2016, rossi2021} than corrected in 
the observations.
With this forward model, we can rigorously analyze galaxy clustering on non-linear scales 
and with higher-order statistics. 

To infer the cosmological parameters, our approach does not require sampling
the posterior using an assumed analytic likelihood. 
We instead use simulation-based inference~\citep[SBI; see][for a review]{cranmer2020}.
SBI, also known as ``likelihood-free inference'', enables accurate Bayesian inference 
using forward models~\citep[\emph{e.g.}][]{papamakarios2017, alsing2019, 
jeffrey2021, tortorelli2021}.
Moreover, they leverage neural density estimation from machine
learning~\citep[\emph{e.g.}][]{germain2015, papamakarios2017} to more 
efficiently infer the posterior without sampling or making strong assumptions on the 
functional form of the likelihood.  

In this work, we apply \simbig~to the CMASS galaxy sample observed by the Sloan 
Digital Sky Survey SDSS-III Baryon Oscillation Spectroscopic 
Survey~\citep[BOSS;][]{eisenstein2011, dawson2013}.
With the main goal of demonstrating the accuracy and potential of \simbig, we use 
the power spectrum as our summary statistic.
We present the cosmological constraints inferred from our
analysis and compare them to previous constraints in the literature.
In an accompanying paper~\citep[][hereafter H22b]{simbig_mocha}, we present our 
forward model in further detail and the mock challenge that we conduct to 
rigorously validate the accuracy of \simbig~cosmological constraints. 

%% file: sbi.tex
\begin{figure*}[]
    \includegraphics[width=17.8cm]{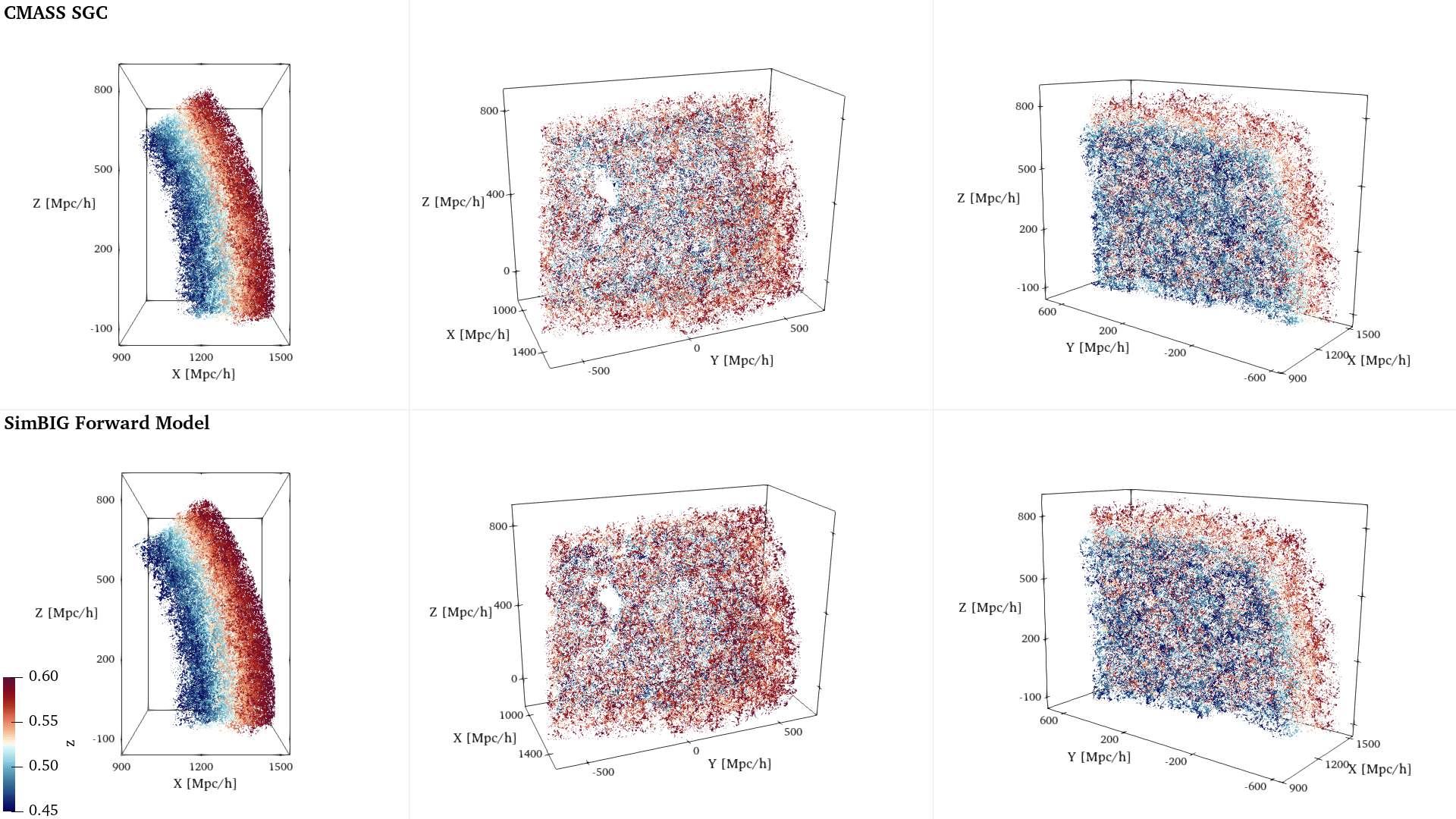}
    \centering
    \caption{
        The \simbig~forward model produces simulated galaxy samples with the same
        survey geometry and observational systematics as the observed BOSS CMASS SGC 
        galaxy sample. 
        We present the 3D distribution of the galaxies from three different viewing 
        angles. 
        The colormap represents the redshift of the galaxies. 
        In the top set of panels,  we present the distribution of galaxies in the CMASS sample. 
        In the bottom, we present the distribution of a simulated galaxy sample, generated from our forward model.  
        The \simbig~galaxy samples are constructed from {\sc Quijote} $N$-body dark
        matter simulations using an HOD model that populates dark matter halos
        identified using the {\sc Rockstar} algorithm. 
        The 3D distributions illustrate that our forward model is able to  generate 
        galaxy distributions that are difficult to statistically distinguish from 
        observations.
        For more comparisons of the 3D distributions, we refer readers to 
        \href{https://youtube.com/playlist?list=PLQk8Faa2x0twK3fgs55ednnHD2vbIzo4z}{\faGlobe}.
    }
    \label{fig:fm}
    \centering
\end{figure*}

\section*{Simulation-Based Inference of Galaxies \simbig} \label{sec:lfi}

Modern cosmological analyses use Bayesian inference to constrain the posterior
distribution $p(\btheta\given \bfi{x})$ of cosmological parameters, $\btheta$, 
given observation $\bfi{x}$. 
In standard galaxy clustering analyses, the posterior is evaluated using Bayes'
rule. 
The likelihood is assumed to have a Gaussian functional form and evaluated using 
an analytic PT model.

SBI offers an alternative that requires no assumptions on the form of the
likelihood. 
SBI only requires a forward model, \emph{i.e.} a simulation to
generate mock observations $\bfi{x}'$ given parameters $\btheta'$. 
It uses a training dataset of simulated pairs $(\btheta', \bfi{x}')$ to estimate 
the posterior. 
SBI has already been successfully applied to a wide range of inference problems
in astronomy and cosmology~\citep{cameron2012, weyant2013, hahn2017b,
alsing2019, huppenkothen2021, jeffrey2021, zhang2021, hahn2022a}.

In this work, we utilize SBI based on neural density estimation, where a neural
network $q$ with parameters $\bphi$ is trained to estimate 
$p(\btheta \given \bfi{x}) \approx q_{\bphi}(\btheta \given \bfi{x})$. 
In particular, we use ``normalizing flow'' models that are capable 
of accurately estimating complex distributions~\citep{tabak2010, tabak2013}.
Below, we briefly describe our forward model and SBI framework. 

\subsection{Forward Model} \label{simbig} 
SBI requires a forward model that is capable of generating mock observations
which are statistically indistinguishable from the observations. 
We start with high-resolution $N$-body simulations from the {\sc Quijote}
suite~\citep{villaescusa-navarro2020}.
These simulations follow the evolution of $1024^3$ cold dark matter (CDM)
particles in a volume of $(1\,h^{-1}{\rm Gpc})^3$ from $z=127$ to $z=0.5$
using the TreePM {\sc Gadget-III} code. 
They accurately model the clustering of matter down to non-linear scales beyond 
$k = 0.5\,h/{\rm Mpc}$~\citep{villaescusa-navarro2020}.

To model the galaxy distribution, we identify gravitationally bound dark
matter halos and populate them with galaxies using a flexible halo occupation 
framework. 
We identify halos using the {\sc Rockstar} phase-space-based halo 
finder~\citep{behroozi2013a}, which accurately determines the location of halos 
and resolves their substructure~\citep{knebe2011}.
We then populate the halos using Halo Occupation Distribution (HOD) models 
that provide a statistical prescription for populating halos with galaxies based 
on halo properties such as their mass and concentration. 
In this work, we use a state-of-the-art HOD model that supplements the standard
\cite{zheng2007} model with assembly, concentration, and velocity biases. 
The extra features of our HOD model add additional flexibility that recent
works suggest may be necessary to describe galaxy
clustering~\citep[\emph{e.g.}][]{zentner2016, vakili2019, hadzhiyska2021}.

Once we have our galaxy distribution in the simulation box, we apply survey realism. 
We remap the box to a cuboid \citep{carlson2010} and then cut out
the detailed survey geometry of the BOSS CMASS SGC sample (see Materials and Methods). 
This includes masking for bright stars, centerpost, bad field, and collision 
priority~\citep{ross2012, dawson2013, ross2017}.
We apply fiber collisions by first identifying all pairs of galaxies within an
angular scale of $62^{\prime\prime}$ then, for 60\% of the pairs, removing one
of the galaxies from the sample. 
Lastly, we trim the forward modeled galaxy catalog to match the $0.45 < z < 0.6$
redshift range and angular range of the observations. 

\begin{figure}
\centering
    \includegraphics[width=7cm]{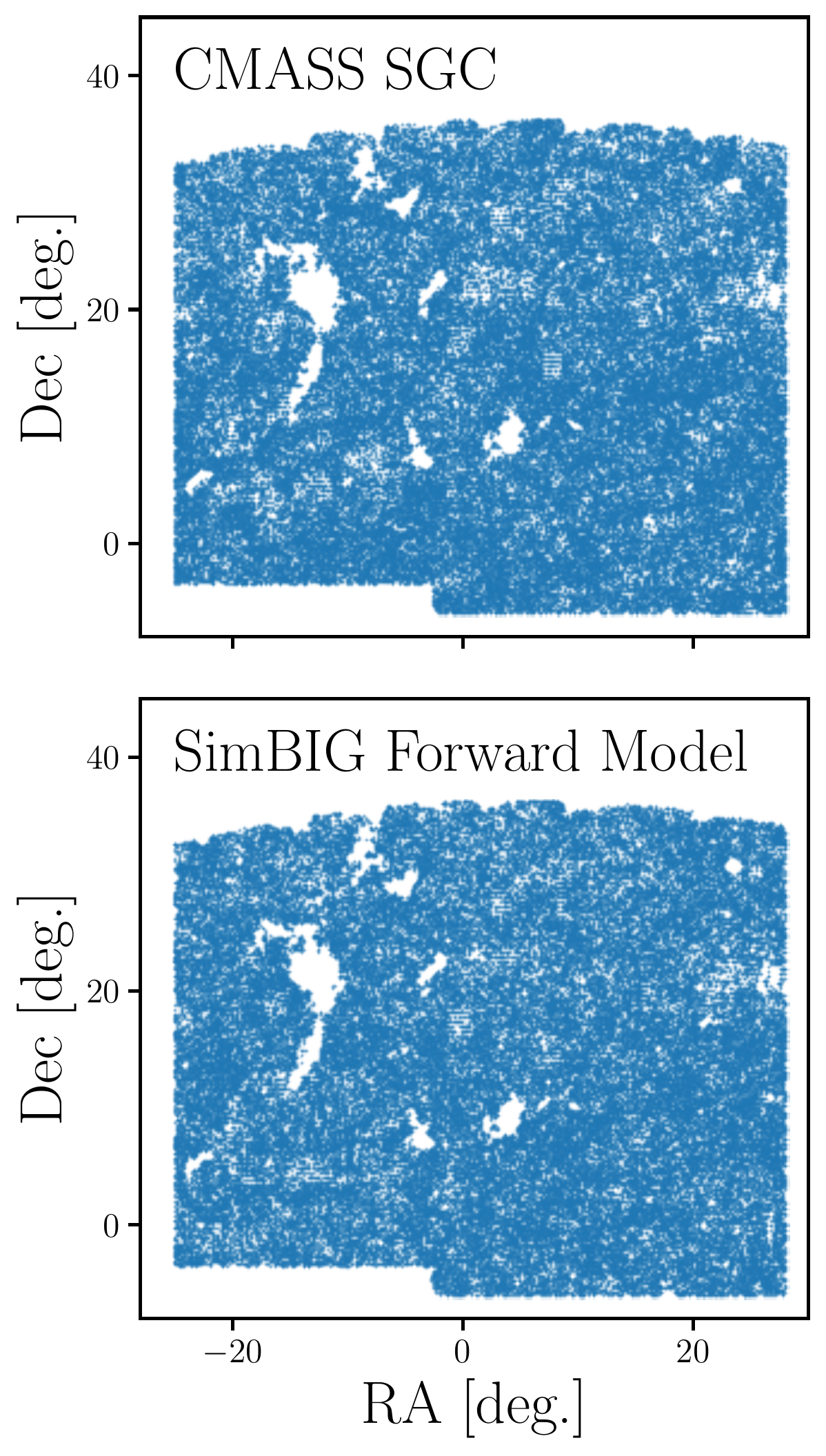}
    \caption{
        Angular distribution of galaxies from the CMASS sample (top) and a galaxy 
        sample generated using the \simbig~forward model (bottom).
        Comparison of the angular distributions highlights the detailed 
        CMASS angular selection that we include in our forward model to 
        account for  observational systematics. 
    }
    \label{fig:fm_demo_ang}
\centering
\end{figure}

In total, our forward model has 14 parameters.
5 $\Lambda$CDM cosmological parameters, $\Omega_m, \Omega_b, h, n_s, \sigma_8$,
that determine the matter distribution and 9 HOD parameters that determine the
connection between galaxies and halos. 
For further details on our forward model, we refer readers to H22b. 
In the bottom panels of Fig.~\ref{fig:fm}, we present the three-dimensional spatial 
distribution of galaxies in our forward  model. 
We present the angular distribution of galaxies in our forward  model in 
Fig.~\ref{fig:fm_demo_ang}. 
The forward model accurately reproduces the survey geometry and angular
footprint of the observed BOSS sample. 
For additional comparisons of the 3D distributions of galaxies in CMASS and our 
forward model, we refer readers to
\href{https://youtube.com/playlist?list=PLQk8Faa2x0twK3fgs55ednnHD2vbIzo4z}{\faGlobe}
\footnote{\url{https://youtube.com/playlist?list=PLQk8Faa2x0twK3fgs55ednnHD2vbIzo4z}}.

\begin{figure*}
    \centering
    \includegraphics[width=14cm]{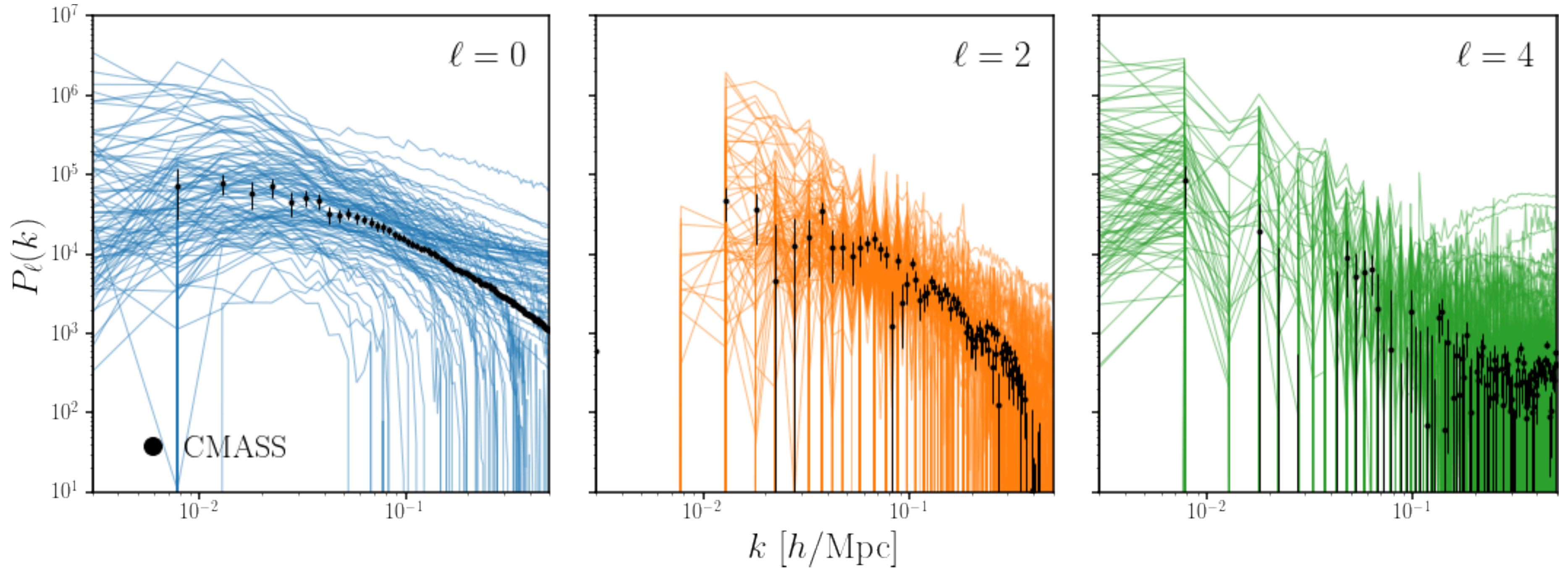}
    \caption{
    Power spectrum, $P_\ell(k)$, measured from the simulated galaxy
    catalogs constructed using the \simbig~forward model.
    We present $P_\ell(k)$ of 100 out of the total 20,000 catalogs for clarity.
    In each of the panels, we plot the monopole, quadrupole, and hexadecapole 
    of the power spectrum ($\ell = 0, 2, 4$).
    For reference, we include $P_\ell(k)$ measured from the BOSS CMASS SGC
    galaxy sample (black) with uncertainties estimated from H22b simulations. 
    $P_\ell$ is the most commonly used summary statistic of galaxy distribution
    that measures the two-point clustering. 
    We use $P_\ell$ in this work to showcase and validate the \simbig~framework 
    and make detailed comparisons to previous works in the literature. 
    The $P_\ell$ of the \simbig~catalogs encompass the BOSS $P_\ell$
    and, thus, provide a sufficiently broad dataset to conduct SBI. 
    }
    \label{fig:plk}
    \centering
\end{figure*}

\subsection{Training Dataset for Simulation-Based Inference}
Using our forward model, we construct 20,000 simulated galaxy catalogs. 
They are constructed from 2,000 {\sc Quijote} $N$-body simulations with 10
different sets of HOD parameters, sampled from a broad prior. 
The $N$-body simulations are arranged in a Latin Hypercube configuration, 
which therefore imposes uniform priors on the cosmological parameters that
conservatively encompasses the {\em Planck} cosmological
constraints~\citep{planckcollaboration2018}. 

In principle, \simbig~can be directly applied to the full galaxy catalog if the
forward model is capable of accurately modeling observations at all scales. 
Even with $N$-body simulations, however, this is not the case due to limitations
on mass and time resolution and inadequacies of halo occupation models.
Instead, we use summary statistics of the galaxy sample, where we can
impose cuts, \emph{e.g.,} based on physical scales, to which our forward model is
accurate. 
Since the primary goal of this work is to present and demonstrate the
\simbig~framework, we use the most commonly used summary statistic: the galaxy
power spectrum multipole, $P_\ell(k)$. 
We also include the average galaxy number density of the sample, $\bar{n}_g$.

In this work, we use the redshift-space galaxy power spectrum
monopole, quadrupole, and hexadecapole  ($\ell=0, 2$, and $4$).
We measure $P_0$, $P_2$, and $P_4$  for each of the simulated galaxy catalogs 
using the \cite{hand2017a} algorithm. 
In this work, we impose a conservative $k < k_{\rm max} = 0.5\,h/{\rm Mpc}$ limit 
on the $P_\ell$, based on the convergence of matter clustering of the {\sc Quijote}
simulations (see H22b for further details). 
We also measure the power spectrum for the BOSS CMASS-SGC galaxy sample with 
the same algorithm.
For the observed $\hat{P}_\ell(k)$ we include systematics weights for 
redshift failures, stellar density, and seeing conditions, which are effects
not included in our forward model but shown to be successfully 
accounted for using the weights~\citep{ross2012, anderson2014}. 

By using $P_\ell$, we can compare the constraints inferred using \simbig~with
previous constraints~\citep[\emph{e.g.}][]{ivanov2020, kobayashi2021} as
further validation of \simbig. 
To be further consistent with previous analyses, we include a nuisance
parameter, $A_{\rm shot}$, that is typically included to account for residual
shot noise 
contribution~\citep[\emph{e.g.}][]{beutler2017, ivanov2020, kobayashi2021}.
In Fig.~\ref{fig:plk}, we present $P_\ell(k)$ of our forward modeled
galaxy catalogs.
We randomly select 100 out of the total 20,000 power spectra for clarity. 
The left, center, and right panels present the monopole, quadrupole, and 
hexadecapole.
These $P_\ell$ measurements and $\bar{n}_g$ serve as the training dataset 
$\{(\btheta', \bfi{x}')\}$  for our SBI posterior estimation using normalizing 
flows. 

\subsection{Simulation-Based Inference with Normalizing Flows}
Normalizing flow models use an invertible bijective transformation, 
$f: \bfi{z} \mapsto \btheta$, to map a complex target distribution to a simple base
distribution, $\pi(\bfi{z})$, that is fast to evaluate. 
For SBI, the target distribution is the posterior, $p(\btheta | \bfi{x})$ while
$\pi(\bfi{z})$ is typically a multivariate Gaussian. 
The transformation $f$ must be invertible and have 
a tractable Jacobian so that we can evaluate the target distribution from
$\pi(\bfi{z})$ by change of variables.
Since $\pi(\bfi{z})$ is easy to sample and evaluate, we can also easily sample and 
evaluate the target distribution.
A neural network with parameters $\bphi$ is trained to obtain $f$.

Among the various normalizing flow-based neural density estimators now available 
in the literature, we use a Masked Autoregressive
Flow~\citep[MAF;][]{papamakarios2017}.
MAF combines normalizing flows with an autoregressive design~\citep{uria2016},
which is well-suited for estimating conditional probability distributions such as
a posterior. 
A MAF model is built by stacking multiple Masked Autoencoder for Distribution
Estimation~\citep[MADE;][]{germain2015} models so that it has the autoregressive
structure of MADE models but with additional flexibility to describe complex 
probability distributions. 
We use the MAF implementation of the \texttt{sbi} Python package~\citep{greenberg2019,
tejero-cantero2020}.

In training, our goal is to determine the parameters, $\bphi$, of our normalizing flow,
$q_{\bphi}(\btheta \given \bfi{x})$, so that it accurately estimates the posterior,
$p(\btheta \given \bfi{x})$. 
We can formulate this into an optimization problem of minimizing the Kullback-Leibler (KL)
divergence between $p(\btheta, \bfi{x}) = p(\btheta\given \bfi{x}) p(\bfi{x})$ and 
$q_{\bphi}(\btheta\given\bfi{x}) p(\bfi{x})$, which measures the difference between the two 
distributions. 
\begin{flalign}
    \min_{\bphi} D_{\rm KL}\bigl(p(\btheta, \bfi{x})\,&\parallel\,q_{\bphi}(\btheta \given \bfi{x}) p(\bfi{x}) \bigl) &&\nonumber \\
    &= \min_{\bphi} \int p(\btheta,\bfi{x})
    \log\frac{p(\btheta\given \bfi{x})}{q_{\bphi}(\btheta \given \bfi{x})}\,{\rm d}\btheta {\rm d}\bfi{x}&&\\
    &\approx \min_{\bphi} \sum\limits_i \log p(\btheta_i \given \bfi{x}_i) - \log q_\phi(\btheta_i \given \bfi{x}_i) \label{eq:kl}\\ 
    &\approx \min_{\bphi} \sum\limits_i - \log q_\phi(\btheta_i \given \bfi{x}_i) \\ 
    &\approx \max_{\bphi} \sum\limits_i \log q_\phi(\btheta_i \given \bfi{x}_i)
    \label{eq:maxkl}.
\end{flalign}
Eq.~\ref{eq:kl} follows from the fact that the training dataset 
$\{(\btheta', \bfi{x}')\}$ is constructed by sampling from $p(\btheta,\bfi{x})$ with our
forward model. 

We split the training data into a training and validation set with a 90/10 split, 
then maximize Eq.~\ref{eq:maxkl} over the training set.
We use the {\sc Adam} optimizer~\citep{kingma2017} with a learning rate of
$5\times10^{-4}$. 
We prevent overfitting by stopping the training when the log-likelihood 
(Eq.~\ref{eq:maxkl}) evaluated on the validation set fails to increase after 20 epochs. 
We determine the architecture of our normalizing flow through experimentation. 
Our final trained model has 6 MADE blocks, each with 9 hidden layers and 186 hidden units.
For further details on the training procedure, we refer readers to H22b. 
Once trained, we estimate the posterior of our 5 cosmological, 9 HOD parameters, and 
1 nuisance parameter.
for the BOSS CMASS SGC $\hat{P_\ell}$ and $\hat{\bar{n}}_g$, by sampling our normalizing 
flow $q_{\bphi}(\btheta\given\hat{\bfi{x}})$.

%% file: results.tex
\begin{figure*}
    \centering
    \includegraphics[width=17.8cm]{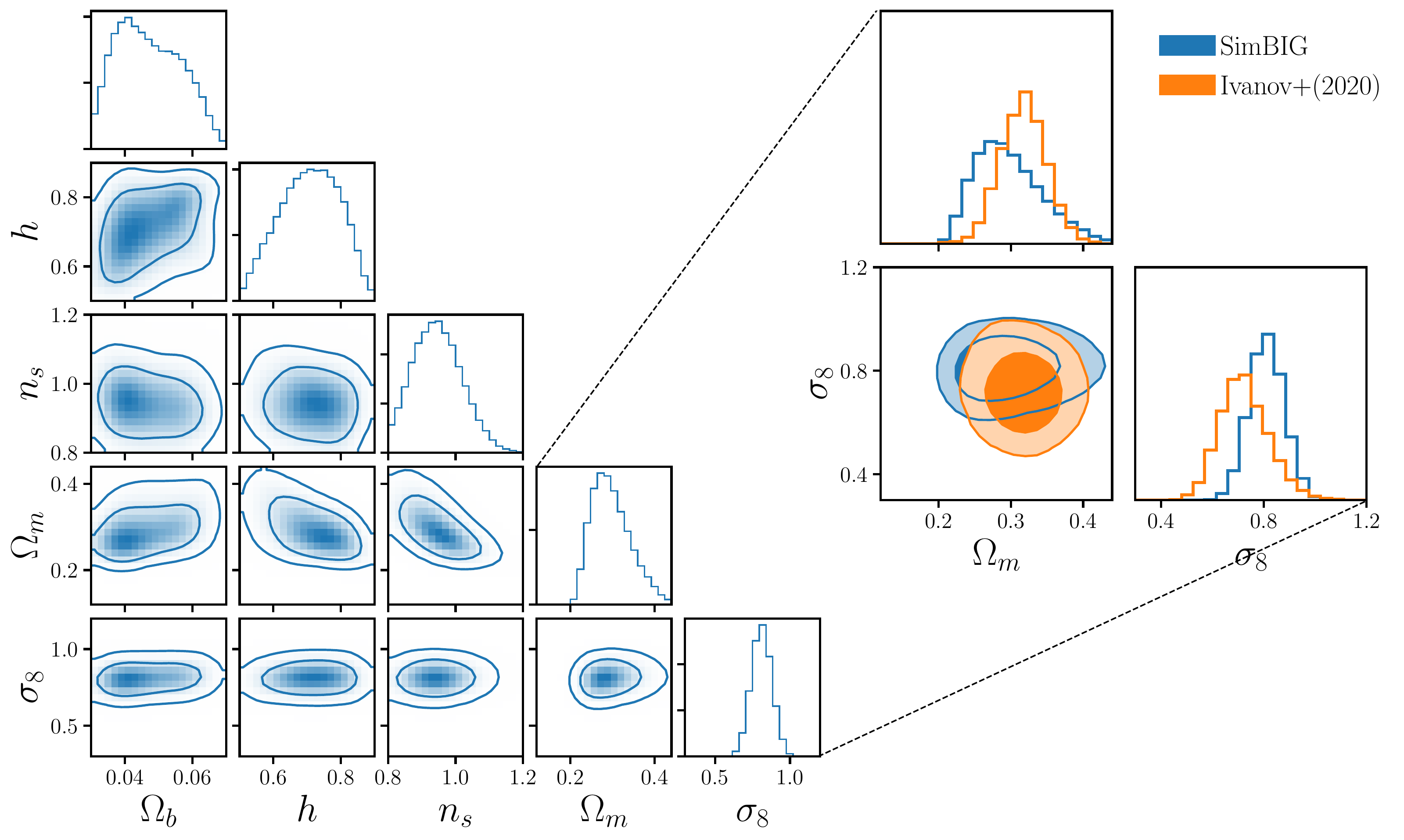}
        \caption{
        {\em Left}: 
        Posterior of cosmological parameters inferred from $P_\ell$ using
        \simbig. 
        In the diagonal panels we present the marginalized 1D posterior of each
        parameter.  
        The other panels present the 2D posteriors that illustrate the degeneracies
        between two parameters.
        The contours mark the 68 and 95 percentiles.
        We accurately analyze $P_\ell$ down to non-linear regimes, 
        $k_{\rm max} = 0.5\,h/{\rm Mpc}$, by using a simulation-based forward model 
        that includes observational systematics. 
        {\em Right}: 
        We focus on the posteriors of ($\Omega_m$, $\sigma_8$), the parameters that 
        can be most significantly constrained by galaxy clustering alone.  
        We derive
        $\Omega_m = 0.292^{+0.055}_{-0.040}$ 
        and 
        $\sigma_8 = 0.812^{+0.067}_{-0.068}$. 
        Our $\sigma_8$ constraints are 27\% tighter than the \cite{ivanov2020} 
        $k_{\rm max} = 0.25\,h/{\rm Mpc}$ PT constraint (orange).
        }
    \label{fig:post}
    \centering
\end{figure*}

\section{Results} \label{results}
We present the posterior distribution of the $\Lambda$CDM cosmological 
parameters, $\Omega_m, \Omega_b, h, n_s, \sigma_8$, inferred from 
$P_\ell(k)$ using \simbig~in Fig.~\ref{fig:post}. 
The posterior is inferred from the BOSS $P_\ell(k)$ down to 
$k_{\rm max} = 0.5\,h/{\rm Mpc}$.
The diagonal panels present the marginalized one-dimensional posteriors 
for each parameter.
The other panels present marginalized two-dimensional posteriors of 
different parameter pairs that highlight parameter degeneracies. 
We mark the 68 and 95 percentiles of the posteriors with the contours. 
We infer the posterior of HOD and nuisance parameters; however, we do
not include them in the figure for clarity. 
Among the cosmological parameters, the \simbig~posterior significantly 
constrains $\Omega_m$ and $\sigma_8$. 
This is consistent with previous works that relied on priors from 
Big Bang nucleosynthesis or cosmic microwave background (CMB) 
experiments for the other parameters, $\Omega_b$ and $n_s$
\citep[\emph{e.g.}][]{ivanov2020, kobayashi2021}.
We infer 
$\Omega_m = 0.292^{+0.055}_{-0.040}$ 
and 
$\sigma_8 = 0.812^{+0.067}_{-0.068}$. 

In the accompanying paper H22b, we present the validation of the
\simbig~posterior using a suite of 1,500 test simulations. 
We construct the test suite using different forward models than the one 
used for our training data.  
They are constructed using different $N$-body simulations, halo finders, 
and HOD models. 
This is to ensure that the cosmological constraints we derive are
independent of the choices and assumptions made in our forward model.  

For validation, we conduct a mock challenge where we infer posteriors of 
the cosmological parameters for each of the test simulations.
Since we know the true cosmological parameter values of the test simulations, 
we can access the accuracy and precision of the inferred posteriors. 
H22b reveals that \simbig~produces unbiased posteriors.
On the other hand, the posteriors are conservative, \emph{i.e.} they are
broader than the true posterior.
This is due to the limited number of $N$-body simulations used to construct 
our training dataset, which makes the estimate of the KL divergence 
(Eq.~\ref{eq:kl}) noisy.
Our constraint on $\Omega_m$ is particularly conservative.  
Additional $N$-body simulations would significantly improve the precision 
of our posteriors. 

Despite the fact that they are conservative, the $\sigma_8$ posterior from 
\simbig~is significantly more precise than constraints from previous works.
\cite{ivanov2020} analyzed the $P_\ell$ of the BOSS CMASS galaxy sample using 
the PT approach with an analytic  model based on effective field theory. 
For the CMASS SGC sample, with uniform priors on the cosmological parameters, and 
with $k_{\rm max} = 0.25\,h/{\rm Mpc}$, \cite{ivanov2020} inferred 
$\sigma_8 = 0.719^{+0.100}_{-0.085}$. 
With \simbig, we improve $\sigma_8$ constraints by 27\% over the standard galaxy
clustering analysis. 
We emphasize that this improvement is roughly equivalent to analyzing a 
galaxy survey $\sim$60\% larger than the original survey using PT.

Recently, \cite{kobayashi2021} also analyzed the $P_\ell$ of BOSS CMASS 
sample but using a theoretical model based on a halo power spectrum emulator. 
Instead of using a galaxy bias scheme used by PT to connect 
the galaxy and matter distributions, \cite{kobayashi2021} used halo power
spectra predicted by an emulator and a halo occupation framework, similar to 
the HOD model in our forward model. 
We note that while the halo power spectrum emulator is trained using simulations, 
the approach in \cite{kobayashi2021}  does not forward model observational 
systematics.
They also make the same assumptions on the form of the likelihood as PT analyses 
for their inference.
For the CMASS SGC sample, with uniform priors on all cosmological parameters,
and with $k_{\rm max} = 0.25\,h/{\rm Mpc}$, \cite{kobayashi2021} inferred 
$\sigma_8 = 0.790^{+0.083}_{-0.072}$. 
The \cite{kobayashi2021} constraints are tighter than the \cite{ivanov2020} PT
constraints because the halo occupation model provides a more compact framework for
modeling galaxies.
Nevertheless, with \simbig, we improve on their $\sigma_8$ constraints by 13\%. 

\simbig~produces significantly tighter constraints on $\sigma_8$ because
we are able to accurately extract cosmological information available on 
small, non-linear, scales.
With our forward modeling approach, we can accurately model non-linear 
clustering and robustly account for observational systematics down to
$k_{\rm max} = 0.5\,h/{\rm Mpc}$. 
In both \cite{ivanov2020} and \cite{kobayashi2021}, they restrict their
analysis to $k_{\rm max} < 0.25\,h/{\rm Mpc}$ due to the limitations of their 
analyses on smaller scales. 

\begin{figure}
\centering
    \includegraphics[width=8.7cm]{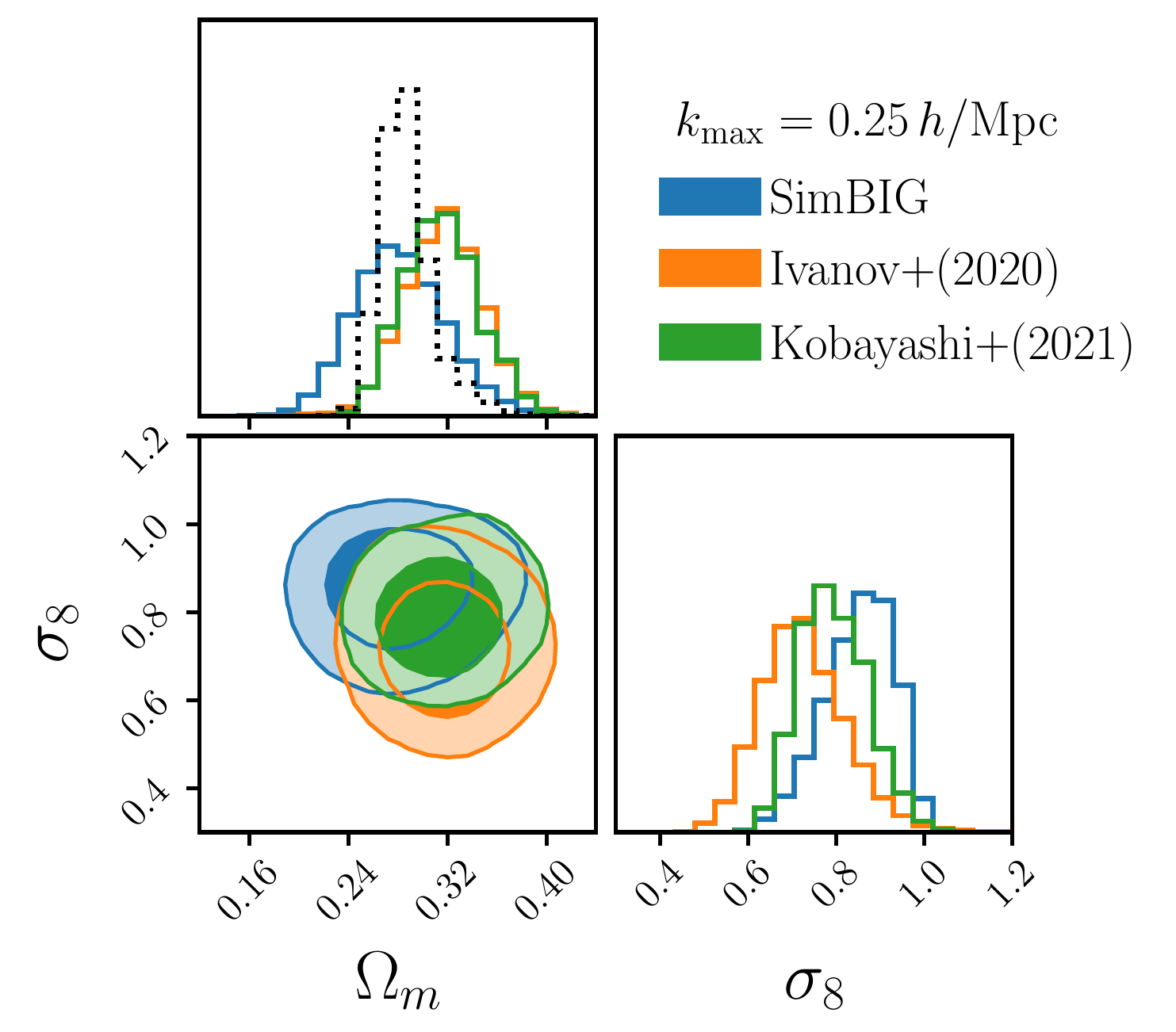}
    \caption{
    Comparison of the ($\Omega_m$, $\sigma_8$) posteriors inferred from the 
    $P_\ell$ CMASS-SGC analysis for $k_{\rm max} = 0.25\,h/{\rm Mpc}$
    from \simbig~(blue), the \cite{ivanov2020} PT approach (orange), and the
    \cite{kobayashi2021} emulator approach (green). 
    The contours represent the 68 and 95 percentiles. 
    We find overall good agreement among the posteriors. 
    For $\Omega_m$, \simbig~infers consistent but broader posterior due to
    the fact that we use a limited number of simulations.  
    We estimate the expected $\Omega_m$ constraints with more simulations 
    using posterior ``recalibration'' (black dotted). 
    For $\sigma_8$, both \simbig~and \cite{kobayashi2021} derive tighter 
    constraints than \cite{ivanov2020} due to the fact that we use halo 
    occupation instead of a galaxy bias prescription.
    Meanwhile, the precision of our $\sigma_8$ constraint is in 
    excellent  agreement with \cite{kobayashi2021}, which uses a similar 
    halo occupation model. 
    The consistency of the $k_{\rm max} = 0.25\,h/{\rm Mpc}$ posteriors demonstrate 
    that the improvements in the $k_{\rm max} = 0.5\,h/{\rm Mpc}$ constraints 
    come from additional cosmological information in the non-linear regime that
    \simbig~can robustly extract. 
    }
    \label{fig:post_comp}
\centering
\end{figure}

To further verify that our improvement comes from constraining power at 
$k > 0.25\,h/{\rm Mpc}$, we analyze $P_\ell$ to
$k_{\rm max} = 0.25\,h/{\rm Mpc}$ using \simbig. 
In Fig.~\ref{fig:post_comp}, we present the \simbig~$k_{\rm max} = 0.25\,h/{\rm Mpc}$
posterior (blue) along with the posteriors from \citep[][orange]{ivanov2020} 
and \citep[][green]{kobayashi2021}.  
We focus our comparison on $\Omega_m$ and $\sigma_8$, the cosmological parameters
that can be most competitively constrained with galaxy clustering alone.
The contours again represent the 68 and 95 percentiles.
We find overall good agreement among the posteriors.  
All of the posteriors are statistically consistent with each other. 
For $\sigma_8$, our $k_{\rm max} = 0.25\,h/{\rm Mpc}$ places a 
$\sigma_8 = 0.861^{+0.070}_{-0.091}$ constraint. 
This is significantly broader than our $k_{\rm max} = 0.5\,h/{\rm Mpc}$
constraint, which demonstrates that the constraining power is in fact
from non-linear scales.
Furthermore, the precision of the $k_{\rm max} = 0.25\,h/{\rm Mpc}$ 
\simbig~constraint is in excellent agreement with the \cite{kobayashi2021} 
constraint.
This serves as further validation of \simbig, since \cite{kobayashi2021} uses a 
similar halo occupation framework to model the power spectrum.

For $\Omega_m$, we infer broader posteriors than \cite{ivanov2020} and \cite{kobayashi2021}.
As we discuss in H22b, this is due to the fact that the \simbig~normalizing 
flow is  trained using a limited number of simulations. 
We use 20,000 forward modeled simulations; however, they are constructed from
2000 $N$-body simulations with different values of cosmological parameters. 
To estimate the expected improvement in the $\Omega_m$ constraints, we use the 
posterior ``re-calibration'' procedure from \cite{dey2022}. 
The re-calibration uses the posteriors inferred for the test simulations 
and their true parameter values. 
We calculate the local probability integral transform~\cite{zhao2021}, 
a diagnostic of the inferred posteriors, and use this quantity to derive 
a weighting scheme that corrects the posteriors so that it matches the
true posterior of the test simulations. 

The re-calibration uses test simulations, so we do not use it for inference. 
However, it provides a bound for the \simbig~constraints if we were to have 
sufficient training simulations. 
The re-calibrated posterior constrains $\Omega_m = 0.284^{+0.021}_{-0.017}$. 
For reference, we mark the re-calibrated $\Omega_m$ constraint (black dotted) in 
Fig.~\ref{fig:post_comp}. 
The re-calibrated $\Omega_m$ is in good agreement with both the \cite{ivanov2020} 
and \cite{kobayashi2021} constraints.
It is significantly tighter than the original \simbig~constraint and illustrates
that additional training simulations would significantly improve the precision 
of the \simbig~$\Omega_m$ constraints. 

Based on our $k_{\rm max} = 0.5\,h/{\rm Mpc}$ posterior, we infer  
$S_8 = \sigma_8 \sqrt{\Omega_m/0.3} = 0.802^{+0.102}_{-0.092}$ 
(and $0.797^{+0.078}_{-0.076}$ for our re-calibrated posterior).  
Multiple recent large-scale structure studies have reported a ``$S_8$ tension'' with 
constraints from the \cite{planckcollaboration2018} CMB analysis.
They find significantly lower values of $S_8$ than {\em Planck}~\citep{cacciato2013, 
mandelbaum2013, leauthaud2017, hikage2019, asgari2021, krolewski2021, amon2022, lange2022}.
PT analyses of BOSS also infer relatively low values of $S_8$.
\cite{ivanov2020}, for instance, infers $S_8 = 0.737^{+0.110}_{-0.092}$.
This $S_8$ tension between constraints from large-scale structure and CMB
analyses has motivated a number of works to explore modifications of the 
standard $\Lambda$CDM cosmological 
model~\citep[\emph{e.g.}][]{meerburg2014, chudaykin2018, divalentino2020, abellan2022}.
We do not find a significant $S_8$ tension with the {\em Planck} 
constraints~\citep{planckcollaboration2018}.
However, given the statistical precision of our $S_8$ constraint, we refrain from 
more detailed comparison and discussion.

%% file: conclusion.tex
\section{Conclusions} 
We present \simbig, a forward modeling framework for analyzing
galaxy clustering using SBI.
As a demonstration of the framework, we apply it to the BOSS CMASS SGC,
a galaxy sample  at $z\sim 0.5$. 
We analyze the galaxy power spectrum multipoles ($P_\ell$), the most
commonly used summary statistic of the galaxy spatial distribution, to 
showcase and validate the \simbig~framework.

\simbig~utilizes a full forward model of the CMASS sample, unlike 
standard approaches that use analytic models of the summary statistic.
The forward model is based on high-resolution {\sc Quijote}
$N$-body simulations that can accurately model the matter distribution 
on small scales.
It uses halo modeling and a state-of-the-art HOD model with assembly, 
concentration, and velocity biases that provide a flexible mapping 
between the matter and galaxy distributions. 
The forward model also includes realistic observational systematics 
such as survey geometry and fiber collisions. 
With this forward modeling approach, we can leverage the predictive
power of simulations to analyze small, non-linear, scales as well as  
higher-order clustering.
It also provides a framework to account for systematics for any summary 
statistic. 

Using the forward model, we construct 20,000 simulated CMASS-like 
samples that span a wide range of cosmological  and HOD parameters. 
We measure $P_\ell$ and $\bar{n}_g$ for each of these samples and use 
the measurements as the training dataset for SBI. 
To estimate the posterior, we use neural density estimation based on
normalizing flows.
Using the training dataset, we train our normalizing flows by minimizing
the KL divergence between its posterior estimate and the true posterior. 
Once trained, we apply our normalizing flow to the observed summary statistics
to infer the posterior of 5 cosmological, 9 HOD, and 1 nuisance parameter.

Focusing on the cosmological parameters, we derive significant constraints
on: 
$\Omega_m =0.316^{+0.040}_{-0.036}$ and 
$\sigma_8 = 0.668^{+0.0324}_{-0.0284}$. 
Our $\sigma_8$ constraints are 27\% tighter than the \cite{ivanov2020} 
constraints using a standard PT approach on the same galaxy sample. 
This improvement is roughly equivalent to increasing the volume of the
galaxy survey by $\sim$60\% for a standard PT analysis. 
The \simbig~constraints are inferred from $P_\ell$ out to 
$k_{\rm max} = 0.5\,h/{\rm Mpc}$ while the PT constraints are limited to
$k_{\rm max} = 0.25\,h/{\rm Mpc}$
The improvement is driven by the additional cosmological information on 
non-linear scales that \simbig~can robustly extract.

We also infer the posterior using \simbig~from $P_\ell$ with 
$k_{\rm max} = 0.25\,h/{\rm Mpc}$ and compare it to posteriors in the literature. 
In particular, the \simbig~$\sigma_8$ constraint for $k_{\rm max} = 0.25\,h/{\rm Mpc}$ 
are in excellent agreement with the constraint from the recent halo model based 
emulator analysis of \cite{kobayashi2021}.
Since they use a similar halo occupation framework as \simbig, 
this comparison firmly verifies the robustness of our constraints. 
The comparison also confirms that the improvement in 
\simbig~$k_{\rm max} = 0.5\,h/{\rm Mpc}$ constraints come from the non-linear 
regime. 
In the accompanying H22b, we present additional tests of \simbig~through a 
mock challenge.
The tests use a suite of 1,500 test simulations constructed with different
forward models to demonstrate that \simbig~produces
unbiased cosmological constraints. 
H22b also presents further details on our forward model and discusses 
posterior constraints on HOD parameters. 

\simbig~can also extract higher-order cosmological information. 
Standard galaxy clustering analyses primarily focus on two-point clustering
statistics. 
Analyses of higher-order statistics have been limited to, \emph{e.g.} the 
bispectrum~\citep{gil-marin2017, philcox2021, damico2022}, and even these
analyses extract only limited cosmological information beyond linear scales.  
In subsequent work, we will use \simbig~to analyse the BOSS CMASS galaxies 
using higher-order statistics (the bispectrum) and non-standard observables 
that contain additional cosmological information: \emph{e.g.}  marked 
power spectrum, skew spectra, void probability functions, and wavelet-
scattering-like statistics. 
We will also apply \simbig~to analyze field-level summary statistics that 
capture all the information in the galaxy field using convolutional 
and graph neural networks.

\simbig~can also be extended to upcoming spectroscopic galaxy surveys observed 
using DESI, PFS, {\em Euclid},  and Roman will probe unprecedented cosmic volumes 
over  the next decade.
They will produce the largest and most detailed three-dimensional maps of 
galaxies in the Universe. 
These surveys are already expected to provide the most precise constraints 
on cosmological parameters using standard analyses.
\simbig~can further exploit the statistical power of these surveys to place 
even tighter constraints on cosmological parameters  and produce the most
stringent tests of the standard $\Lambda$CDM cosmological 
model and beyond.

%% file: main.bbl
\begin{thebibliography}{10}

\bibitem{desicollaboration2016}
D Collaboration, et~al., The {{DESI Experiment Part I}}:
  {{Science}},{{Targeting}}, and {{Survey Design}}.
\newblock {\em\protect\JournalTitle{arXiv:1611.00036 [astro-ph]}} (2016).

\bibitem{desicollaboration2016a}
D Collaboration, et~al., The {{DESI Experiment Part II}}: {{Instrument
  Design}}.
\newblock {\em\protect\JournalTitle{arXiv:1611.00037 [astro-ph]}} (2016).

\bibitem{abareshi2022}
B Abareshi, et~al., Overview of the {{Instrumentation}} for the {{Dark Energy
  Spectroscopic Instrument}} (2022).

\bibitem{takada2014}
M Takada, et~al., Extragalactic science, cosmology, and {{Galactic}}
  archaeology with the {{Subaru Prime Focus Spectrograph}}.
\newblock {\em\protect\JournalTitle{Publications of the Astronomical Society of
  Japan}} \textbf{66}, R1 (2014).

\bibitem{tamura2016}
N Tamura, et~al., Prime {{Focus Spectrograph}} ({{PFS}}) for the {{Subaru}}
  telescope: Overview, recent progress, and future perspectives in {\em
  Ground-Based and {{Airborne Instrumentation}} for {{Astronomy VI}}}.
\newblock ({eprint: arXiv:1608.01075}), Vol.{} 9908, p. 99081M (2016).

\bibitem{laureijs2011}
R Laureijs, et~al., Euclid {{Definition Study Report}}.
\newblock {\em\protect\JournalTitle{arXiv e-prints}} p. arXiv:1110.3193 (2011).

\bibitem{spergel2015}
D Spergel, et~al., Wide-{{Field InfrarRed Survey Telescope-Astrophysics Focused
  Telescope Assets WFIRST-AFTA}} 2015 {{Report}} (2015).

\bibitem{wang2022a}
Y Wang, et~al., The {{High Latitude Spectroscopic Survey}} on the {{Nancy Grace
  Roman Space Telescope}}.
\newblock {\em\protect\JournalTitle{The Astrophysical Journal}} \textbf{928}, 1
  (2022).

\bibitem{beutler2017}
F Beutler, et~al., The clustering of galaxies in the completed {{SDSS-III
  Baryon Oscillation Spectroscopic Survey}}: Anisotropic galaxy clustering in
  {{Fourier}} space.
\newblock {\em\protect\JournalTitle{Monthly Notices of the Royal Astronomical
  Society}} \textbf{466}, 2242--2260 (2017).

\bibitem{ivanov2020}
MM Ivanov, M Simonovi{\'c}, M Zaldarriaga, Cosmological parameters from the
  {{BOSS}} galaxy power spectrum.
\newblock {\em\protect\JournalTitle{Journal of Cosmology and Astroparticle
  Physics}} \textbf{2020}, 042 (2020).

\bibitem{kobayashi2021}
Y Kobayashi, T Nishimichi, M Takada, H Miyatake, Full-shape cosmology analysis
  of {{SDSS-III BOSS}} galaxy power spectrum using emulator-based halo model: A
  \$5\textbackslash\%\$ determination of \$\textbackslash sigma\_8\$.
\newblock {\em\protect\JournalTitle{arXiv:2110.06969 [astro-ph]}} (2021).

\bibitem{bernardeau2002}
F Bernardeau, S Colombi, E Gaztanaga, R Scoccimarro, Large-{{Scale Structure}}
  of the {{Universe}} and {{Cosmological Perturbation Theory}}.
\newblock {\em\protect\JournalTitle{Physics Reports}} \textbf{367}, 1--248
  (2002).

\bibitem{desjacques2016}
V Desjacques, D Jeong, F Schmidt, Large-{{Scale Galaxy Bias}}.
\newblock {\em\protect\JournalTitle{arXiv:1611.09787 [astro-ph, physics:gr-qc,
  physics:hep-ph]}} (2016).

\bibitem{philcox2021}
OHE Philcox, MM Ivanov, The {{BOSS DR12 Full-Shape Cosmology}}:
  \$\textbackslash{{Lambda}}\${{CDM Constraints}} from the {{Large-Scale Galaxy
  Power Spectrum}} and {{Bispectrum Monopole}}.
\newblock {\em\protect\JournalTitle{arXiv:2112.04515 [astro-ph,
  physics:hep-ex]}} (2021).

\bibitem{damico2022}
G D'Amico, Y Donath, M Lewandowski, L Senatore, P Zhang, The {{BOSS}}
  bispectrum analysis at one loop from the {{Effective Field Theory}} of
  {{Large-Scale Structure}} (2022).

\bibitem{naidoo2022}
K Naidoo, E Massara, O Lahav, Cosmology and neutrino mass with the {{Minimum
  Spanning Tree}}.
\newblock {\em\protect\JournalTitle{Monthly Notices of the Royal Astronomical
  Society}} (2022).

\bibitem{ross2012}
AJ Ross, et~al., The clustering of galaxies in the {{SDSS-III Baryon
  Oscillation Spectroscopic Survey}}: Analysis of potential systematics.
\newblock {\em\protect\JournalTitle{Monthly Notices of the Royal Astronomical
  Society}} \textbf{424}, 564--590 (2012).

\bibitem{ross2017}
AJ Ross, et~al., The clustering of galaxies in the completed {{SDSS-III Baryon
  Oscillation Spectroscopic Survey}}: Observational systematics and baryon
  acoustic oscillations in the correlation function.
\newblock {\em\protect\JournalTitle{Monthly Notices of the Royal Astronomical
  Society}} \textbf{464}, 1168--1191 (2017).

\bibitem{guo2012}
H Guo, I Zehavi, Z Zheng, A {{New Method}} to {{Correct}} for {{Fiber
  Collisions}} in {{Galaxy Two-point Statistics}}.
\newblock {\em\protect\JournalTitle{The Astrophysical Journal}} \textbf{756},
  127 (2012).

\bibitem{hahn2017a}
C Hahn, R Scoccimarro, MR Blanton, JL Tinker, SA {Rodr{\'i}guez-Torres}, The
  {{Effect}} of {{Fiber Collisions}} on the {{Galaxy Power Spectrum
  Multipoles}}.
\newblock {\em\protect\JournalTitle{Monthly Notices of the Royal Astronomical
  Society}} \textbf{467}, 1940--1956 (2017).

\bibitem{bianchi2018}
D Bianchi, et~al., Unbiased clustering estimates with the {{DESI}} fibre
  assignment.
\newblock {\em\protect\JournalTitle{Monthly Notices of the Royal Astronomical
  Society}} \textbf{481}, 2338--2348 (2018).

\bibitem{zehavi2002}
I Zehavi, et~al., Galaxy {{Clustering}} in {{Early Sloan Digital Sky Survey
  Redshift Data}}.
\newblock {\em\protect\JournalTitle{The Astrophysical Journal}} \textbf{571},
  172--190 (2002).

\bibitem{anderson2014}
L Anderson, et~al., The clustering of galaxies in the {{SDSS-III Baryon
  Oscillation Spectroscopic Survey}}: Baryon acoustic oscillations in the
  {{Data Releases}} 10 and 11 {{Galaxy}} samples.
\newblock {\em\protect\JournalTitle{Monthly Notices of the Royal Astronomical
  Society}} \textbf{441}, 24--62 (2014).

\bibitem{hahn2020}
C Hahn, F {Villaescusa-Navarro}, E Castorina, R Scoccimarro, Constraining
  {{M$\nu$}} with the bispectrum. {{Part I}}. {{Breaking}} parameter
  degeneracies.
\newblock {\em\protect\JournalTitle{Journal of Cosmology and Astroparticle
  Physics}} \textbf{03}, 040 (2020).

\bibitem{hahn2021a}
C Hahn, F {Villaescusa-Navarro}, Constraining {{M$\nu$}} with the bispectrum.
  {{Part II}}. {{The}} information content of the galaxy bispectrum monopole.
\newblock {\em\protect\JournalTitle{Journal of Cosmology and Astroparticle
  Physics}} \textbf{2021}, 029 (2021).

\bibitem{massara2020}
E Massara, F {Villaescusa-Navarro}, S Ho, N Dalal, DN Spergel, Using the
  {{Marked Power Spectrum}} to {{Detect}} the {{Signature}} of {{Neutrinos}} in
  {{Large-Scale Structure}}.
\newblock {\em\protect\JournalTitle{arXiv:2001.11024 [astro-ph]}} (2020).

\bibitem{massara2022}
E Massara, et~al., Cosmological {{Information}} in the {{Marked Power
  Spectrum}} of the {{Galaxy Field}} (2022).

\bibitem{wang2022}
Y Wang, et~al., Extracting high-order cosmological information in galaxy
  surveys with power spectra (2022).

\bibitem{hou2022}
J {Hou}, A {Moradinezhad Dizgah}, C {Hahn}, E {Massara}, {Cosmological
  Information in Skew Spectra of Biased Tracers in Redshift Space}.
\newblock {\em\protect\JournalTitle{arXiv e-prints}} p. arXiv:2210.12743
  (2022).

\bibitem{eickenberg2022}
M Eickenberg, et~al., Wavelet {{Moments}} for {{Cosmological Parameter
  Estimation}} (2022).

\bibitem{rodriguez-torres2016}
SA {Rodr{\'i}guez-Torres}, et~al., The clustering of galaxies in the {{SDSS-III
  Baryon Oscillation Spectroscopic Survey}}: Modelling the clustering and halo
  occupation distribution of {{BOSS CMASS}} galaxies in the {{Final Data
  Release}}.
\newblock {\em\protect\JournalTitle{Monthly Notices of the Royal Astronomical
  Society}} \textbf{460}, 1173--1187 (2016).

\bibitem{rossi2021}
G Rossi, et~al., The {{Completed SDSS-IV Extended Baryon Oscillation
  Spectroscopic Survey}}: {{N-body Mock Challenge}} for {{Galaxy Clustering
  Measurements}}.
\newblock {\em\protect\JournalTitle{Monthly Notices of the Royal Astronomical
  Society}} \textbf{505}, 377--407 (2021).

\bibitem{cranmer2020}
K Cranmer, J Brehmer, G Louppe, The frontier of simulation-based inference.
\newblock {\em\protect\JournalTitle{Proceedings of the National Academy of
  Sciences}} \textbf{117}, 30055--30062 (2020).

\bibitem{papamakarios2017}
G Papamakarios, T Pavlakou, I Murray, Masked {{Autoregressive Flow}} for
  {{Density Estimation}}.
\newblock {\em\protect\JournalTitle{arXiv e-prints}} \textbf{1705},
  arXiv:1705.07057 (2017).

\bibitem{alsing2019}
J Alsing, T Charnock, S Feeney, B Wandelt, Fast likelihood-free cosmology with
  neural density estimators and active learning.
\newblock {\em\protect\JournalTitle{Monthly Notices of the Royal Astronomical
  Society}} \textbf{488}, 4440--4458 (2019).

\bibitem{jeffrey2021}
N Jeffrey, J Alsing, F Lanusse, Likelihood-free inference with neural
  compression of {{DES SV}} weak lensing map statistics.
\newblock {\em\protect\JournalTitle{Monthly Notices of the Royal Astronomical
  Society}} \textbf{501}, 954--969 (2021).

\bibitem{tortorelli2021}
L Tortorelli, et~al., The {{PAU Survey}}: {{Measurement}} of {{Narrow-band}}
  galaxy properties with {{Approximate Bayesian Computation}}.
\newblock {\em\protect\JournalTitle{arXiv:2106.02651 [astro-ph]}} (2021).

\bibitem{germain2015}
M Germain, K Gregor, I Murray, H Larochelle, {{MADE}}: {{Masked Autoencoder}}
  for {{Distribution Estimation}}.
\newblock {\em\protect\JournalTitle{Proceedings of the 32nd International
  Conference on Machine Learning}} \textbf{37}, 881--889 (2015).

\bibitem{eisenstein2011}
DJ Eisenstein, et~al., {{SDSS-III}}: {{Massive Spectroscopic Surveys}} of the
  {{Distant Universe}}, the {{Milky Way}}, and {{Extra-Solar Planetary
  Systems}}.
\newblock {\em\protect\JournalTitle{The Astronomical Journal}} \textbf{142}, 72
  (2011).

\bibitem{dawson2013}
KS Dawson, et~al., The {{Baryon Oscillation Spectroscopic Survey}} of
  {{SDSS-III}}.
\newblock {\em\protect\JournalTitle{The Astronomical Journal}} \textbf{145}, 10
  (2013).

\bibitem{simbig_mocha}
C {Hahn}, et~al., {SimBIG: Mock Challenge for a Forward Modeling Approach to
  Galaxy Clustering}.
\newblock {\em\protect\JournalTitle{arXiv}} (2022).

\bibitem{cameron2012}
E Cameron, AN Pettitt, Approximate {{Bayesian Computation}} for astronomical
  model analysis: A case study in galaxy demographics and morphological
  transformation at high redshift.
\newblock {\em\protect\JournalTitle{Monthly Notices of the Royal Astronomical
  Society}} \textbf{425}, 44--65 (2012).

\bibitem{weyant2013}
A Weyant, C Schafer, WM {Wood-Vasey}, Likelihood-free {{Cosmological
  Inference}} with {{Type Ia Supernovae}}: {{Approximate Bayesian Computation}}
  for a {{Complete Treatment}} of {{Uncertainty}}.
\newblock {\em\protect\JournalTitle{The Astrophysical Journal}} \textbf{764},
  116 (2013).

\bibitem{hahn2017b}
C Hahn, et~al., Approximate {{Bayesian Computation}} in {{Large Scale
  Structure}}: Constraining the galaxy-halo connection.
\newblock {\em\protect\JournalTitle{Monthly Notices of the Royal Astronomical
  Society}} \textbf{469}, 2791--2805 (2017).

\bibitem{huppenkothen2021}
D Huppenkothen, M Bachetti, Accurate {{X-ray Timing}} in the {{Presence}} of
  {{Systematic Biases With Simulation-Based Inference}} (2021).

\bibitem{zhang2021}
K {Zhang}, et~al., {Real-time Likelihood-free Inference of Roman Binary
  Microlensing Events with Amortized Neural Posterior Estimation}.
\newblock {\em\protect\JournalTitle{Astronomical Journal}} \textbf{161}, 262
  (2021).

\bibitem{hahn2022a}
C Hahn, P Melchior, Accelerated {{Bayesian SED Modeling}} using {{Amortized
  Neural Posterior Estimation}} (2022).

\bibitem{tabak2010}
EG Tabak, E {Vanden-Eijnden}, Density estimation by dual ascent of the
  log-likelihood.
\newblock {\em\protect\JournalTitle{Communications in Mathematical Sciences}}
  \textbf{8}, 217--233 (2010).

\bibitem{tabak2013}
EG Tabak, CV Turner, A {{Family}} of {{Nonparametric Density Estimation
  Algorithms}}.
\newblock {\em\protect\JournalTitle{Communications on Pure and Applied
  Mathematics}} \textbf{66}, 145--164 (2013).

\bibitem{villaescusa-navarro2020}
F {Villaescusa-Navarro}, et~al., The {{Quijote Simulations}}.
\newblock {\em\protect\JournalTitle{The Astrophysical Journal Supplement
  Series}} \textbf{250}, 2 (2020).

\bibitem{behroozi2013a}
PS Behroozi, RH Wechsler, HY Wu, The {{ROCKSTAR Phase-space Temporal Halo
  Finder}} and the {{Velocity Offsets}} of {{Cluster Cores}}.
\newblock {\em\protect\JournalTitle{The Astrophysical Journal}} \textbf{762},
  109 (2013).

\bibitem{knebe2011}
A Knebe, et~al., Haloes gone {{MAD}}: {{The Halo-Finder Comparison Project}}.
\newblock {\em\protect\JournalTitle{Monthly Notices of the Royal Astronomical
  Society}} \textbf{415}, 2293--2318 (2011).

\bibitem{zheng2007}
Z Zheng, AL Coil, I Zehavi, Galaxy {{Evolution}} from {{Halo Occupation
  Distribution Modeling}} of {{DEEP2}} and {{SDSS Galaxy Clustering}}.
\newblock {\em\protect\JournalTitle{The Astrophysical Journal}} \textbf{667},
  760--779 (2007).

\bibitem{zentner2016}
AR Zentner, A Hearin, FC van~den Bosch, JU Lange, A Villarreal, Constraints on
  {{Assembly Bias}} from {{Galaxy Clustering}}.
\newblock {\em\protect\JournalTitle{arXiv:1606.07817 [astro-ph]}} (2016).

\bibitem{vakili2019}
M Vakili, C Hahn, How {{Are Galaxies Assigned}} to {{Halos}}? {{Searching}} for
  {{Assembly Bias}} in the {{SDSS Galaxy Clustering}}.
\newblock {\em\protect\JournalTitle{The Astrophysical Journal}} \textbf{872},
  115 (2019).

\bibitem{hadzhiyska2021}
B Hadzhiyska, et~al., Galaxy assembly bias and large-scale distribution: A
  comparison between {{IllustrisTNG}} and a semi-analytic model.
\newblock {\em\protect\JournalTitle{Monthly Notices of the Royal Astronomical
  Society}} \textbf{508}, 698--718 (2021).

\bibitem{carlson2010}
J Carlson, M White, Embedding realistic surveys in simulations through volume
  remapping.
\newblock {\em\protect\JournalTitle{The Astrophysical Journal Supplement
  Series}} \textbf{190}, 311--314 (2010).

\bibitem{planckcollaboration2018}
{Planck Collaboration}, et~al., Planck 2018 results. {{VI}}. {{Cosmological}}
  parameters.
\newblock {\em\protect\JournalTitle{arXiv:1807.06209 [astro-ph]}} (2018).

\bibitem{hand2017a}
N Hand, Y Li, Z Slepian, U Seljak, An optimal {{FFT-based}} anisotropic power
  spectrum estimator.
\newblock {\em\protect\JournalTitle{Journal of Cosmology and Astro-Particle
  Physics}} \textbf{07}, 002 (2017).

\bibitem{uria2016}
B Uria, MA C{\^o}t{\'e}, K Gregor, I Murray, H Larochelle, Neural
  {{Autoregressive Distribution Estimation}}.
\newblock {\em\protect\JournalTitle{arXiv:1605.02226 [cs]}} (2016).

\bibitem{greenberg2019}
DS Greenberg, M Nonnenmacher, JH Macke, Automatic {{Posterior Transformation}}
  for {{Likelihood-Free Inference}} (2019).

\bibitem{tejero-cantero2020}
A {Tejero-Cantero}, et~al., Sbi: {{A}} toolkit for simulation-based inference.
\newblock {\em\protect\JournalTitle{Journal of Open Source Software}}
  \textbf{5}, 2505 (2020).

\bibitem{kingma2017}
DP Kingma, J Ba, Adam: {{A Method}} for {{Stochastic Optimization}}.
\newblock {\em\protect\JournalTitle{arXiv:1412.6980 [cs]}} (2017).

\bibitem{dey2022}
B Dey, et~al., Calibrated {{Predictive Distributions}} via {{Diagnostics}} for
  {{Conditional Coverage}} (2022).

\bibitem{zhao2021}
D Zhao, N Dalmasso, R Izbicki, AB Lee, Diagnostics for conditional density
  models and {{Bayesian}} inference algorithms in {\em Proceedings of the
  {{Thirty-Seventh Conference}} on {{Uncertainty}} in {{Artificial
  Intelligence}}}.
\newblock ({PMLR}), pp. 1830--1840 (2021).

\bibitem{cacciato2013}
M Cacciato, FC {van den Bosch}, S More, H Mo, X Yang, Cosmological constraints
  from a combination of galaxy clustering and lensing - {{III}}.
  {{Application}} to {{SDSS}} data.
\newblock {\em\protect\JournalTitle{Monthly Notices of the Royal Astronomical
  Society}} \textbf{430}, 767--786 (2013).

\bibitem{mandelbaum2013}
R Mandelbaum, et~al., Cosmological parameter constraints from galaxy-galaxy
  lensing and galaxy clustering with the {{SDSS DR7}}.
\newblock {\em\protect\JournalTitle{Monthly Notices of the Royal Astronomical
  Society}} \textbf{432}, 1544--1575 (2013).

\bibitem{leauthaud2017}
A Leauthaud, et~al., Lensing is low: Cosmology, galaxy formation or new
  physics?
\newblock {\em\protect\JournalTitle{Monthly Notices of the Royal Astronomical
  Society}} \textbf{467}, 3024--3047 (2017).

\bibitem{hikage2019}
C Hikage, et~al., Cosmology from cosmic shear power spectra with {{Subaru Hyper
  Suprime-Cam}} first-year data.
\newblock {\em\protect\JournalTitle{Publications of the Astronomical Society of
  Japan}} \textbf{71}, 43 (2019).

\bibitem{asgari2021}
M Asgari, et~al., {{KiDS-1000}} cosmology: {{Cosmic}} shear constraints and
  comparison between two point statistics.
\newblock {\em\protect\JournalTitle{Astronomy \&amp; Astrophysics, Volume 645,
  id.A104, {$<$}NUMPAGES{$>$}31{$<$}/NUMPAGES{$>$} pp.}} \textbf{645}, A104
  (2021).

\bibitem{krolewski2021}
A Krolewski, S Ferraro, M White, Cosmological constraints from {{unWISE}} and
  {{Planck CMB}} lensing tomography.
\newblock {\em\protect\JournalTitle{Journal of Cosmology and Astroparticle
  Physics}} \textbf{2021}, 028 (2021).

\bibitem{amon2022}
A Amon, et~al., Dark {{Energy Survey Year}} 3 results: {{Cosmology}} from
  cosmic shear and robustness to data calibration.
\newblock {\em\protect\JournalTitle{Physical Review D}} \textbf{105}, 023514
  (2022).

\bibitem{lange2022}
JU Lange, et~al., Five per cent measurements of the growth rate from
  simulation-based modelling of redshift-space clustering in {{BOSS LOWZ}}.
\newblock {\em\protect\JournalTitle{Monthly Notices of the Royal Astronomical
  Society}} \textbf{509}, 1779--1804 (2022).

\bibitem{meerburg2014}
PD Meerburg, Alleviating the tension at low $l$ through axion monodromy.
\newblock {\em\protect\JournalTitle{Physical Review D}} \textbf{90}, 063529
  (2014).

\bibitem{chudaykin2018}
A Chudaykin, D Gorbunov, I Tkachev, Dark matter component decaying after
  recombination: {{Sensitivity}} to baryon acoustic oscillation and redshift
  space distortion probes.
\newblock {\em\protect\JournalTitle{Physical Review D}} \textbf{97}, 083508
  (2018).

\bibitem{divalentino2020}
E Di~Valentino, A Melchiorri, O Mena, S Vagnozzi, Nonminimal dark sector
  physics and cosmological tensions.
\newblock {\em\protect\JournalTitle{Physical Review D}} \textbf{101}, 063502
  (2020).

\bibitem{abellan2022}
GF Abell{\'a}n, R Murgia, V Poulin, J Lavalle, Implications of the {{S8}}
  tension for decaying dark matter with warm decay products.
\newblock {\em\protect\JournalTitle{Physical Review D}} \textbf{105}, 063525
  (2022).

\bibitem{gil-marin2017}
H {Gil-Mar{\'i}n}, et~al., The clustering of galaxies in the {{SDSS-III Baryon
  Oscillation Spectroscopic Survey}}: {{RSD}} measurement from the power
  spectrum and bispectrum of the {{DR12 BOSS}} galaxies.
\newblock {\em\protect\JournalTitle{Monthly Notices of the Royal Astronomical
  Society}} \textbf{465}, 1757--1788 (2017).

\bibitem{reid2016}
B Reid, et~al., {{SDSS-III Baryon Oscillation Spectroscopic Survey Data
  Release}} 12: Galaxy target selection and large-scale structure catalogues.
\newblock {\em\protect\JournalTitle{Monthly Notices of the Royal Astronomical
  Society}} \textbf{455}, 1553--1573 (2016).

\end{thebibliography}
